\documentclass[twocolumn,floatfix,showpacs,prl,superscriptaddress]{revtex4}
\usepackage{mathrsfs}
\usepackage{graphicx,graphics,color,epsfig}
\usepackage{bm}
\usepackage{amsmath}
\usepackage{amssymb}
\usepackage{subfigure}
\usepackage{amsfonts}
\usepackage[colorlinks=true, breaklinks=true, linkcolor=blue, citecolor=blue,urlcolor=blue, pdfborder={0 0 0}]{hyperref}
\providecommand{\U}[1]{\protect\rule{.1in}{.1in}}
\newcommand{\doi}[2]{\href{http://dx.doi.org/#1}{#2}}
\newcommand{\arxiv}[1]{\href{http://arxiv.org/abs/#1}{arXiv:#1}}

\begin{document}
\title{Frustration-Induced Gaplessness in the Frustrated Transverse Ising Ring}
\author{Jian-Jun Dong}
\affiliation{College of Physical Science and Technology, Sichuan University, 610064,
Chengdu, China}
\affiliation{Key Laboratory of High Energy Density Physics and Technology of Ministry of
Education, Sichuan University, Chengdu, 610064, China}
\author{Peng Li}
\email{lipeng@scu.edu.cn}
\affiliation{College of Physical Science and Technology, Sichuan University, 610064,
Chengdu, China}
\affiliation{Key Laboratory of High Energy Density Physics and Technology of Ministry of
Education, Sichuan University, Chengdu, 610064, China}
\author{Qi-Hui Chen}
\affiliation{Key Laboratory of Advanced Technologies of Materials (Ministry of Education of
China). Superconductivity R$\&$D Center (SRDC), Mail Stop 165$\#$, Southwest
Jiaotong University, Chengdu, Sichuan 610031, China}
\date{\today }

\begin{abstract}
New effects in the frustrated transverse Ising ring are predicted. The system
is solved based on a mapping of Pauli spin operators to the Jordan-Wigner
fermions. We group the low-lying energy levels into bands after imposing
appropriate parity constraint, which projects out the
redundant degrees of freedom brought about by the Jordan-Wigner fermions. In
the region of strong antiferromagnetic coupling, we uncover an unusual gapless
phase induced by the ring frustration. We demonstrate that its ground state
exhibits a strong longitudinal spin-spin correlation and possesses a
considerably large entropy of entanglement. The low-lying energy levels evolve
adiabatically in the gapless phase, which facilitates us to work out new
behaviors of density of states, low-temperature correlation functions and
specific heat. We also propose an experimental protocol
for observing this peculiar gapless phase.

\end{abstract}

\maketitle

\begin{figure}[ptb]
\includegraphics[width=0.47\textwidth]{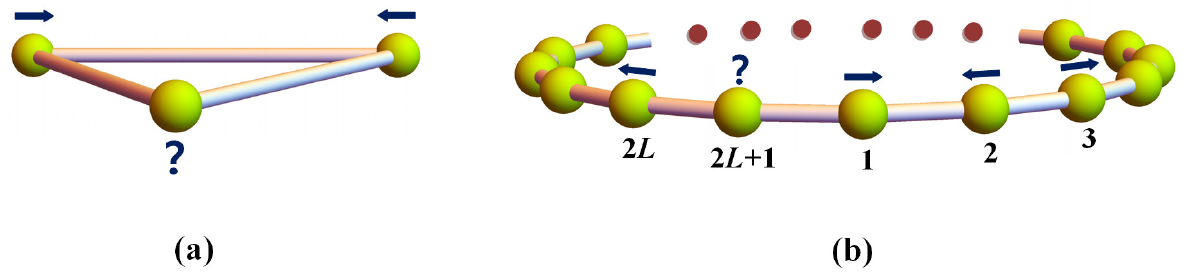}\newline\caption{(Color online)
Periodic ring with odd number of sites: (a)$N=3$, (b)$N=2L+1$ for general
situation. A ring frustration is induced if we consider the Hamiltonian Eq.
(\ref{H1}).}%
\label{fig1}%
\end{figure}

\emph{Introduction.}---Ising chain in transverse field is an excellent
prototype for demonstrating quantum phase transition \cite{Sachdev}. It can be
solved by the Jordan-Wigner transformation, which links the spins and the
Jordan-Wigner fermions through a non-local relation~\cite{J-W}. Its
theoretical properties have been related to real materials since decades ago
\cite{Suzuki}. There is a minor mismatch between the Jordan-Wigner fermions
and the spins in that the transformation depends on boundary condition
\cite{Lieb,Pfeuty}. Nevertheless, the discrepancy in the usual problem is
negligible and the fermion representation can produce reliable results for the
ground state and low excitations. To magnify the mismatch between the fermions
and spins, we work on a frustrated transverse Ising ring (FTIR), i.e. the
frustrated antiferromagnetic Ising model in a transverse field on a periodic
chain with odd number of lattice sites. The model reads
\begin{equation}
H=J\sum_{j=1}^{N}\sigma_{j}^{x}\sigma_{j+1}^{x}-h\sum_{j=1}^{N}\sigma_{j}^{z},
\label{H1}%
\end{equation}
where the number of lattice sites $N$=$2L$+$1$ ($L$=$1$,$2$,$3$,$\ldots$),
$\sigma_{j}^{\alpha}$'s ($\alpha$=$x,z$) are Pauli matrices, the longitudinal
antiferromagnetic coupling $J$$>$$0$, the uniform transverse field $h$$>$ $0$.
A periodic boundary condition $\sigma_{N+1}^{\alpha}$=$\sigma_{1} ^{\alpha} $
is imposed. One can choose the representation that ensures the diagonalization
of $\sigma_{j}^{z}$, i.e. $\sigma_{j}^{z}\left\vert \uparrow_{j}\right\rangle
$=$\left\vert \uparrow_{j}\right\rangle $ and $\sigma_{j}^{z}\left\vert
\downarrow_{j}\right\rangle $=$\left\vert \downarrow_{j}\right\rangle $. With
$N$ being odd, the first term in Eq. (\ref{H1}) possesses a ring frustration,
i.e. one can not accomplish an alternative arrangement of the two eigenstates
of $\sigma_{j}^{x}$, $\left\vert \rightarrow_{j}\right\rangle $=$(\left\vert
\uparrow_{j}\right\rangle $+$\left\vert \downarrow_{j}\right\rangle )/\sqrt
{2}$ and $\left\vert \leftarrow_{j}\right\rangle $=$(\left\vert \uparrow_{j}
\right\rangle -\left\vert \downarrow_{j}\right\rangle )/\sqrt{2}$
\cite{Sachdev}, along the chain as shown in Fig. \ref{fig1}.

Frustration is an important topic in quantum magnetism both theoretically and
experimentally \cite{Liebmann,Diep}. However exactly solvable model is rare.
We found the FTIR exhibits remarkable properties due to the ring frustration
despite its simplicity. We will show the frustration induces a peculiar
gapless phase in the strong antiferromagnetic coupling region in thermodynamic
limit. We analyze significant quantities of the ground state and low-energy
excitations to characterize this gapless phase.

\emph{Jordan-Wigner fermions.}---It is convenient to replace the Pauli
matrices with the raising and lowering operators $\sigma_{j}^{\pm}$
=$(\sigma_{j}^{x}\pm\sigma_{j}^{y})/2$. Then one can replace the spin
operators with Jordan-Wigner fermions as $\sigma_{1}^{+}$=$c_{1}^{\dag}$
and$\quad\sigma_{j}^{+}$=$c_{j}^{\dag}\exp($i$\pi\sum_{l=1}^{j-1}c_{l}^{\dag
}c_{l})$ with $j$=$2$, $3$, $\cdots$, $N$ \cite{J-W}. Let us write down the
Hamiltonian after the Jordan-Wigner transformation \cite{Lieb},
\begin{align}
H  &  =Nh-2h\sum_{j=1}^{N}c_{j}^{\dag}c_{j}+J\sum_{j=1}^{N-1}(c_{j}^{\dag
}-c_{j})(c_{j+1}^{\dag}-c_{j+1})\nonumber\\
&  \quad-J\exp(\text{i}\pi M)(c_{N}^{\dag}-c_{N})(c_{1}^{\dag}+c_{1}).
\label{H2}%
\end{align}
Although the total number of fermions $M$=$\sum_{j=1}^{N}c_{j}^{\dag}c_{j}$
does not conserve, the parity $P$=$\exp($i$\pi M)$=$(-1)^{M}$ does. The parity
of the system remains to be fixed in context of boundary condition. If
neglecting the last term in Eq. (\ref{H2}), one would get a so-called
"c-cycle" problem, but we persist on the original "a-cycle" problem here
\cite{Lieb}. It is obvious the fermion problem must obey the periodic boundary
condition (PBC), $c_{1}^{\dag}$=$c_{N+1}^{\dag}$, if $M$ is an odd number, and
the anti-periodic boundary condition (anti-PBC), $c_{1}^{\dag}$=$-c_{N+1}%
^{\dag}$, if $M$ is an even number \cite{Suzuki}. So there are two routes to
be followed, which can be called \emph{odd channel (o)} and \emph{even channel
(e)} respectively. By introducing a Fourier transformation, $c_{q}$%
=$(1/\sqrt{N})\sum_{j}c_{j}\exp($i$qj)$, one would find the momentum must take
a value in the set $q^{o}$=$\{2m\pi/N|\forall m\}$ for odd channel or $q^{e}%
$=$\{(2m$+$1)\pi/N|\forall m\}$ for even channel. In each channel, there are
$N$ available values, i.e. $m$=$-(N-1)/2$, $\ldots$, $-2$, $-1$, $0$, $1$,
$2$, $\ldots$, $(N-1)/2$. Then one can diagonalize the Hamiltonian as
\begin{equation}
H^{o/e}=\sum_{q\in q^{o/e},q\neq q^{\star}}\omega(q)\left(  2\eta_{q}%
^{\dagger}\eta_{q}-1\right)  +\epsilon(q^{\star})\left(  2c_{q\star}^{\dagger
}c_{q\star}-1\right)  \label{H3}%
\end{equation}
by introducing the Bogoliubov transformation $\eta_{q}$=$u_{q}c_{q}-$
i$v_{q}c_{-q}^{\dagger}$ , where the coefficients satisfy the relations,
$u_{q}^{2}$=$\left(  1\text{+}\epsilon(q)/\omega(q)\right)  /2$, $v_{q}^{2}%
$=$\left(  1-\epsilon(q)/\omega(q)\right)  /2$, $2u_{q}v_{q}$=$\Delta
(q)/\omega(q)$, $\omega(q)$=$\sqrt{\epsilon(q)^{2}+\Delta(q)^{2}}$,
$\epsilon(q)$=$J\cos{q}-h$, and $\Delta(q)$=$J\sin{q}$. Notice we do nothing
on the operator $c_{q\star}$ with $q^{\star}$=$0$ for odd channel and
$q^{\star}$=$\pi$ for even channel. It will be helpful to introduce a
quasi-particle number $Q=\sum_{q\in\left(  q^{o}\cup q^{e}\right)  }n_{q}$,
where we have defined: $n_{0}=c_{0}^{\dagger}c_{0}$ for $q=0$, $n_{\pi}%
=c_{\pi}^{\dagger}c_{\pi}$ for $q=\pi$, and $n_{q}=\eta_{q}^{\dagger}\eta_{q}$
for other $q$. It is easy to confirm that $P$=$(-1)^{M}$=$(-1)^{Q}$. The
degrees of freedom (DOF) of the fermionic Hamiltonian are $2^{N}$ for both
channels, so we get $2^{N+1}$ DOF totally, which is redundantly twice of the
DOF of the original spin Hamiltonian. However, the odd channel requires an odd
parity and the even channel an even parity. This parity constraint will help
us to obliterate exactly the redundant DOF in each channel.

\begin{figure}[ptb]
\includegraphics[width=0.47\textwidth]{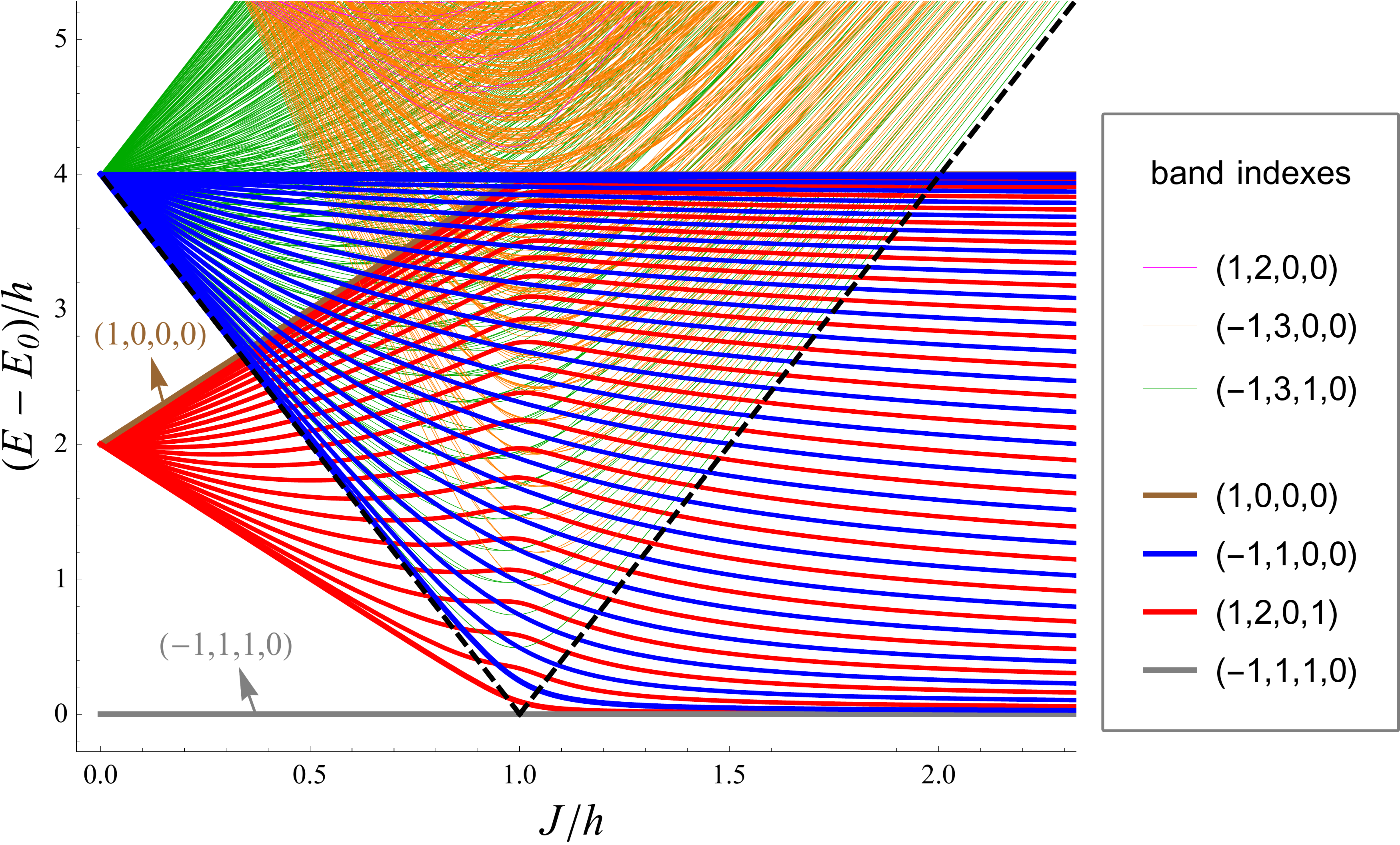}\newline\caption{(Color online)
Bands of low energy levels of the frustrated transverse Ising ring. We take
$N=51$ for a graphical view. The non-degenerate ground state with band indexes
$(P,Q,n_{0},n_{\pi})=(-1,1,1,0)$ is set as the reference. The dashed black
line denotes the lower bound of all bands other than the lowest four ones
listed in the right panel. Its intersecting point at $J/h=1$ is a critical
point in thermodynamic limit, whose right-hand side will become a gapless
region.}%
\label{fig2}%
\end{figure}

\emph{Bands of energy levels.}---Let us seek for the ground state first. Due
to the Bogoliubov transformation, we know the ground state must contain a
BCS-like part. Since there are two channels, we have two pure BCS-like
functions:
\begin{align}
|\phi^{o}\rangle &  =\prod_{\substack{q\in q^{o},q\neq0}}\left(
u_{q}+\text{i}v_{q}c_{q}^{\dag}c_{-q}^{\dag}\right)  |0\rangle,\label{BCSo}\\
|\phi^{e}\rangle &  =\prod_{\substack{q\in q^{e},q\neq\pi}}\left(
u_{q}+\text{i}v_{q}c_{q}^{\dag}c_{-q}^{\dag}\right)  |0\rangle. \label{BCSe}%
\end{align}
It is easy to check that the state $|\phi^{o}\rangle$ does not satisfy the
parity constraint of the odd channel and should be left out. While $|\phi
^{e}\rangle$ is a valid state, however, it possesses too high energy to become
the ground state. The true ground state comes from the odd channel and can be
easily constructed by $|\phi^{o}\rangle$ as
\begin{equation}
|E_{0}\rangle=c_{0}^{\dagger}|\phi^{o}\rangle, \label{E0state}%
\end{equation}
whose energy reads $E_{0}$=$\Lambda^{o}$+$2\omega(0)\theta(J/h-1)$ where
$\Lambda^{o}$=$-\sum_{q\in q^{o}}\omega(q)$ and $\theta(x)$ is a Heaviside
step function. In thermodynamic limit $N\rightarrow\infty$, the ground state
energy per site reads $E_{0}/N$=$\frac{-2}{\pi}\left\vert J-h\right\vert
E\left(  -4Jh/(J-h)^{2}\right)  $+$2(J-h)\theta(J/h-1)/N$, where $E(x)$ is the
complete elliptic integral of the second kind. It is non-analytic at $J/h$
=$1$, because its second derivative in respect of $J/h$ has a logarithmic
divergent peak $\sim(1/\pi)\ln|J/h-1|$, which heralds a critical point. In
fact, the self-duality \cite{Kogut} of FTIR still holds, which ensures this
critical point. One can see this clear by defining new Ising-type operators,
$\tau_{j}^{z}$=$-\sigma_{j}^{x}\sigma_{j+1}^{x}$ and $\tau_{j}^{x}$
=$(-1)^{j}\prod_{k<j}\sigma_{k}^{z}$, to get a dual form of Hamiltonian,
$H$=$-J\sum_{j=1}^{N}\tau_{j}^{z}$+$h\sum_{j=1}^{N-1}\tau_{j}^{x}\tau
_{j+1}^{x}$.

Above the ground state, there is an energy gap $\Delta_{gap}$=$2(h-J)$ to the
first excited states for $J/h$$<$ $1$, which is similar to the usual case
\cite{Sachdev}. While for $J/h$ $>$$1$, the system becomes gapless
surprisingly. This unexpected gapless phase is the main focus of this Letter.
As shown in Fig. \ref{fig2}, there are $2N$ levels involved in the low-energy
properties. All these $2N$ levels can be uniquely designated by $q$ and
constructed by $\eta_{q}^{\dagger}$ in regard of the parity of the channel.
They can be grouped into four bands that are labelled by a set of indexes
$(P,Q,n_{0},n_{\pi})$, because the levels with the same set of indexes are
degenerate at $J$=$0$ and emanate to form a band with $J$ increasing. The
lowest level $|E_{0}\rangle$\ with indexes $(-1,1,1,0)$ and the upper-most
level $|\phi^{e}\rangle$ with indexes $(1,0,0,0)$ are non-degenerate (we shall
label it as $|E_{\pi}\rangle=|\phi^{e}\rangle$). Between them, the levels from
even and odd channels deplete the width of $4h$\ with alternatively increasing
energies. Each level is doubly degenerate. The levels from even channel are
$|E_{q}\rangle$=$\eta_{q}^{\dagger}c_{\pi}^{\dagger}|\phi^{e}\rangle$ ($q\in
q^{e}$, $q\neq\pi$) with indexes ($1$,$2$,$0$,$1$), whose energy reads $E_{q}%
$=$\Lambda^{e}$+$2\omega(q)$ with$\quad\Lambda^{e}$=$-\sum_{q\in q^{e}}%
\omega(q)$. While the ones from odd channel are $|E_{q}\rangle$=$\eta
_{q}^{\dagger}|\phi^{o}\rangle$ ($q\in q^{o}$, $q\neq0$) with indexes
($-1$,$1$,$0$,$0$), whose energy reads $E_{q}$=$\Lambda^{o}$+$2\omega(q)$. We
have $\Lambda$=$\Lambda^{o}$=$\Lambda^{e}$ if $N$ is large enough.

\begin{figure}[ptb]
\includegraphics[width=0.4\textwidth]{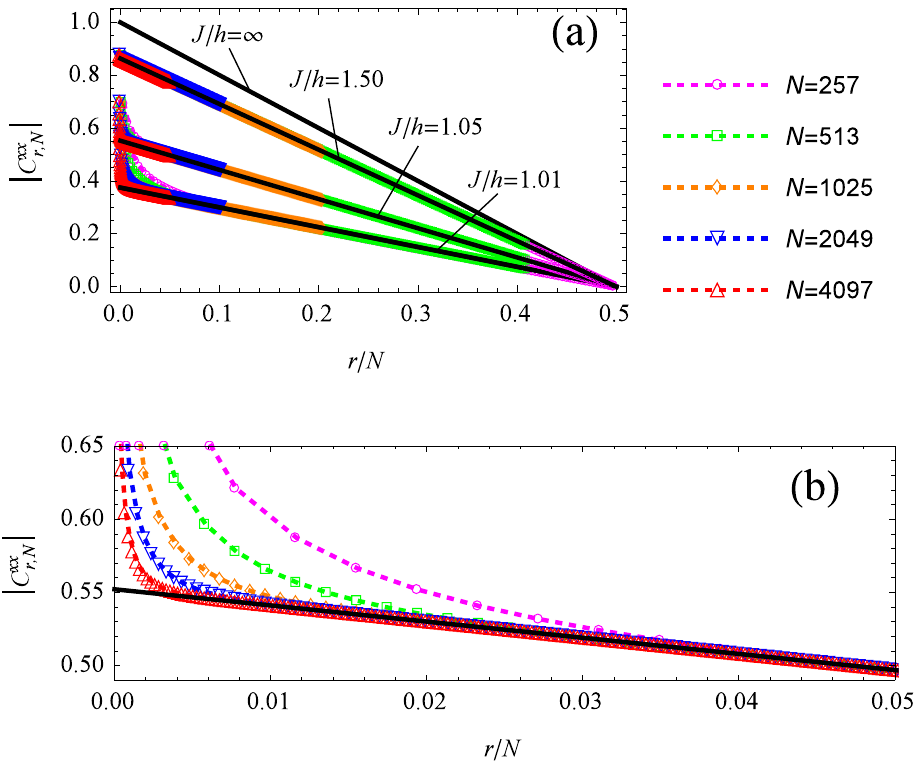}\newline\caption{(Color online)
Longitudinal correlation functions for several selected values of $J/h$ in the
gapless phase. In (a), the colored dingbat data are numerical results from the
determinant Eq. (\ref{Cxxdet}), while the black straight lines are analytical
results from Eq. (\ref{Cxx2}). (b) is a zoom-in plot for the case $J/h=1.05$.}%
\label{fig3}%
\end{figure}

\emph{Correlation function of the ground state.}---The two-point longitudinal
spin-spin correlation function of the ground state is defined as $C_{r,N}%
^{xx}$=$\langle E_{0}|\sigma_{j}^{x}\sigma_{j+r}^{x}|E_{0}\rangle$, where we
notice the result does not depend on $j$. If we introduce the operators,
$A_{j}$=$c_{j}^{\dag}$+$c_{j}$ and $B_{j}$=$c_{j}^{\dag}-c_{j}$, and use the
relations, $A_{j}^{2}$=$1$ and $A_{j}B_{j}$=$\exp(-$i$\pi c_{j}^{\dag}c_{j})$,
we have
\begin{equation}
C_{r,N}^{xx}=\langle\phi^{o}|c_{0}B_{j}A_{j+1}B_{j+1}\ldots B_{j+r-1}%
A_{j+r}c_{0}^{\dag}|\phi^{o}\rangle. \label{Cxx}%
\end{equation}
To evaluate this function thoroughly, we need to make use of the Wick's
theorem and the contractions \cite{Lieb} in respect of $|\phi^{o}\rangle$:
$\langle c_{0}c_{0}^{\dag}\rangle$=$1$, $\langle A_{j}c_{0}^{\dag}\rangle
$=$1/\sqrt{N}$, $\langle B_{j}c_{0}^{\dag}\rangle$=$-1/\sqrt{N}$, $\langle
A_{j}A_{j+r}\rangle$=$\delta_{r,0}$, $\langle B_{j}B_{j+r}\rangle$%
=$-\delta_{r,0}$, and $\langle B_{j}A_{j+r}\rangle$=$D_{r+1}$ with%
\begin{equation}
D_{r}=\frac{1}{N}\sum_{_{\substack{q\in q^{o},q\neq0}}}\exp\left(
-\text{i}qr\right)  D(\operatorname{e}^{\text{i}q})-\frac{1}{N}, \label{Dr}%
\end{equation}
where $D(\operatorname{e}^{\text{i}q})$=$-(J-h\operatorname{e}^{\text{i}%
q})/\omega(q)$. Then we arrive at a Toeplitz determinant%
\begin{equation}
C_{r,N}^{xx}=\left\vert
\begin{array}
[c]{cccc}%
D_{0}+\frac{2}{N} & D_{-1}+\frac{2}{N} & \cdots & D_{-r+1}+\frac{2}{N}\\
D_{1}+\frac{2}{N} & D_{0}+\frac{2}{N} & \cdots & D_{-r+2}+\frac{2}{N}\\
\cdots & \cdots & \cdots & \cdots\\
D_{r-1}+\frac{2}{N} & D_{r-2}+\frac{2}{N} & \cdots & D_{0}+\frac{2}{N}%
\end{array}
\right\vert . \label{Cxxdet}%
\end{equation}
For finite $N$, this determinant can be evaluated numerically. While for
$N\rightarrow\infty$, we need to substitute the sum $\frac{1}{N}%
\sum_{_{\substack{q\in q^{o},q\neq0}}}$ in Eq. (\ref{Dr}) with an integral
${\int}\frac{dq}{2\pi}$ by defining appropriate $D(\operatorname{e}%
^{\text{i}0})$ to evaluate the determinant analytically \cite{McCoy}.

First, let us focus on the most intriguing gapless phase ($J/h$ $>$$1$).
Following the earlier procedure of Wu \cite{McCoy,McCoy Wu,Wu}, we put forward
a proposition \cite{S1}:

\textit{Proposition: }Consider a Toeplitz determinant%
\begin{equation}
\Theta(r,N,x,\operatorname{e}^{\operatorname*{i}k})=\left\vert
\begin{array}
[c]{cccc}%
\overset{\sim}{D}_{0} & \overset{\sim}{D}_{-1} & \cdots & \overset{\sim
}{D}_{1-r}\\
\overset{\sim}{D}_{1} & \overset{\sim}{D}_{0} & \cdots & \overset{\sim
}{D}_{2-r}\\
\cdots & \cdots & \cdots & \cdots\\
\overset{\sim}{D}_{r-1} & \overset{\sim}{D}_{r-2} & \cdots & \overset{\sim
}{D}_{0}%
\end{array}
\right\vert \label{Theta}%
\end{equation}
with $\overset{\sim}{D}_{n}$=$\int_{-\pi}^{\pi}\frac{dq}{2\pi}%
\,D(\operatorname{e}^{\operatorname*{i}q})\,\operatorname{e}%
^{-\operatorname*{i}qn}+\frac{x}{N}\operatorname{e}^{\operatorname*{i}kn}$. If
the generating function $D(\operatorname{e}^{\operatorname*{i}q})$ and $\ln
D(\operatorname{e}^{\operatorname*{i}q})$\ are continuous on the unit circle
$\left\vert \operatorname{e}^{\operatorname*{i}q}\right\vert =1$, then the
behavior for large $N$ of $\Theta(r,N,x,\operatorname{e}^{\operatorname*{i}%
k})$ is given by ($1\ll r\ll N$)
\begin{equation}
\Theta(r,N,x,\operatorname{e}^{\operatorname*{i}k})=\Delta_{r}(1+\frac
{xr}{ND(\operatorname{e}^{-\operatorname*{i}k})}),
\end{equation}
\ where $\Delta_{r}$=$\mu^{r}\exp(\sum_{n=1}^{\infty}nd_{-n}d_{n})$,$\mu
$=$\exp[\int_{-\pi}^{\pi}\frac{dq}{2\pi}\,\ln D(\operatorname{e}%
^{\operatorname*{i}q})]$, and $d_{n}$=$\int_{-\pi}^{\pi}\frac{dq}{2\pi
}\,\operatorname{e}^{-\operatorname*{i}qn}\ln D(\operatorname{e}%
^{\operatorname*{i}q})$, if the sum $\sum_{n=1}^{\infty}nd_{-n}d_{n}$ is convergent.

By applying this proposition to the gapless phase ($x$=$2$ and $k$=$0$), we
get
\begin{equation}
C_{r,N}^{xx}\approx(-1)^{r}(1-\frac{h^{2}}{J^{2}})^{1/4}(1-\frac{2r}{N})
\label{Cxx2}%
\end{equation}
for large enough $r$ and $N$. This asymptotic behaviour is depicted in Fig. 3,
which is perfectly coincident with the numerical results. For $h=0$, Eq.
(\ref{Cxx2}) becomes rigorous. This surprising result is totally different
from the conventional case without frustration \cite{Sachdev}, and needs a
further elucidation.

While for the gapped phase ($J/h$ $<$ $1$), the Toeplitz determinant recovers
the usual case without frustration \cite{McCoy}. The resulting correlation
function decays exponentially with a finite correlation length $\xi
=-1/\ln(J/h)$ \cite{McCoy1968,Suzuki}. And for the critical point ($J/h$=$1$
and $N\rightarrow\infty$), the correlation function decays algebraically as
$C_{r,\infty}^{xx}\sim(-1)^{r}\alpha/r^{1/4}$ \cite{McCoy}. Our numerical
results are in good agreement with these conclusions, which will not be
presented here since they are not the focus of this Letter.

\begin{figure}[ptb]
\includegraphics[width=0.4\textwidth]{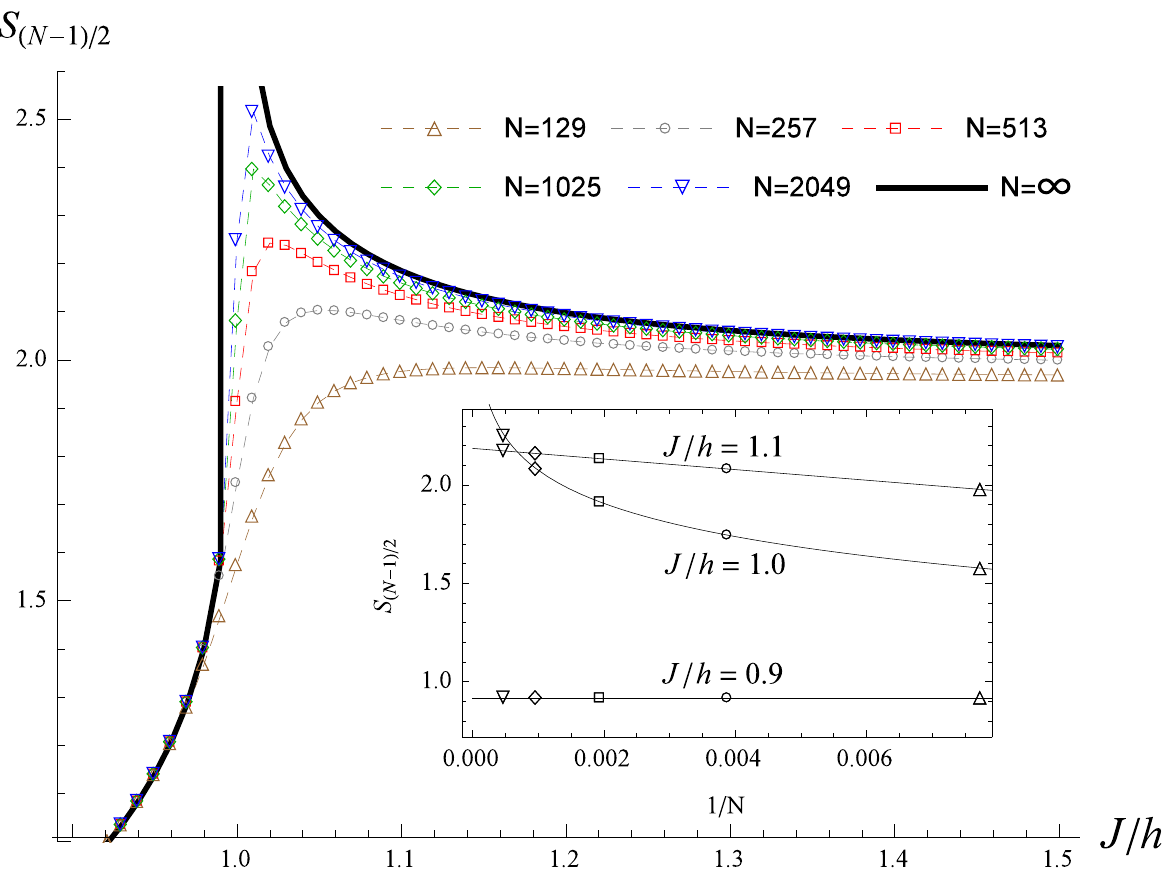}\newline\caption{(Color online)
Entanglement entropy $S_{(N-1)/2}$ as a function of $J/h$ for a sequence of
number of lattice sites $N$. The inset shows examples of finite size scaling
for extrapolating to the thermodynamic limit. At the critical point, the
numerical data fit a divergent behavior, $S_{(N-1)/2}\sim\frac{1}{6}\log_{2}%
N$, coincident with the prediction by CFT \cite{Holzhey}.}%
\label{fig4}%
\end{figure}

\emph{Entanglement entropy of the ground state.}---The entanglement entropy
(EE) is defined as the von Neumann entropy $S_{l}$=$-$tr$(\rho_{l}\log
_{2}\rho_{l})$ of the reduced density matrix $\rho_{l}$=tr$_{N-l}|E_{0}%
\rangle\langle E_{0}|$, where the trace is performed on the spin states of
contiguous sites from $j$=$1$ to $N-l$ \cite{Vidal}. We can evaluate the EE
numerically by utilizing the matrix
\begin{equation}
\Gamma_{l}=\left\vert
\begin{array}
[c]{cccc}%
\Pi_{0} & \Pi_{1} & \cdots & \Pi_{l-1}\\
\Pi_{-1} & \Pi_{0} & \cdots & \Pi_{l-2}\\
\cdots & \cdots & \cdots & \cdots\\
\Pi_{1-l} & \Pi_{2-l} & \cdots & \Pi_{0}%
\end{array}
\right\vert \text{ with }\Pi_{l}=\left\vert
\begin{array}
[c]{cc}%
0 & -g_{l}\\
g_{-l} & 0
\end{array}
\right\vert ,
\end{equation}
where $g_{l}$=$D_{l-1}$+$\frac{2}{N}$. Let $V\in SO(2l)$ denote an orthogonal
matrix that brings $\Gamma_{l}$ into a block diagonal form such that
$\Gamma_{l}^{C}$=$V\Gamma_{l}V^{T}$=$\bigoplus_{m=0}^{l-1}($i$v_{m}\sigma
_{y})$ with $v_{m}\geq0$. Then $S_{l}$ is given by $S_{l}$=$\sum_{m=0}%
^{l-1}H_{2}(\frac{1+v_{m}}{2})$ with $H_{2}(x)$=$-x\log_{2}x-(1-x)\log
_{2}(1-x)$. The numerical results for $l$=$(N-1)/2$ are shown in Fig.
\ref{fig4}. We observe the EE of the gapped phase is small until near the
critical point, where it abruptly tends to become divergent as predicted by
CFT \cite{Holzhey}. While in the gapless phase, we observe $S_{(N-1)/2}$ with
$N\rightarrow\infty$ approaches its minimal value $2$ when $h\rightarrow0$,
where the FTIR reduces to the classical Ising model with the ring frustration,
$H^{\text{Ising}}$=$J\sum_{j=1}^{N}\sigma_{j}^{x}\sigma_{j+1}^{x}$. The EE of
the gapless phase is much larger than that of the well-known GHZ states, whose
EE $S_{(N-1)/2}$ is exactly $\log_{2}2$=$1$ \cite{Vidal}. The $2N$-degenerate
ground states of $H^{\text{Ising}}$ can be denoted by kink states
\cite{Solomon}:
\begin{equation}
\genfrac{\{}{.}{0pt}{}{|K(j),\rightarrow\rangle=|\cdots,\leftarrow
_{j-1},\rightarrow_{j},\rightarrow_{j+1},\leftarrow_{j+2},\cdots
\rangle,}{|K(j),\leftarrow\rangle=|\cdots,\rightarrow_{j-1},\leftarrow
_{j},\leftarrow_{j+1},\rightarrow_{j+2},\cdots\rangle,}
\label{kink}
\end{equation}
where kinks occur between sites $j$ and $j+1$. The kink states break parity
symmetry. One can verify the ground state $|E_{0}\rangle$ with odd parity
evolves adiabatically into a simple sum of the kink states when $h\rightarrow
0$:
\begin{equation}
\lim_{h\rightarrow0}|E_{0}\rangle=\frac{1}{\sqrt{2N}}{\sum\nolimits_{j,\tau}%
}|K(j),\tau\rangle,
\end{equation}
whose EE is exactly $2$ in thermodynamic limit. In fact, all of the degenerate
$2N$ states near $h\sim0$ are just recombinations of the kink states. We
usually take it for granted\ that the classical system $H^{\text{Ising}}$
falls into one of the kink states at zero temperature due to spontaneous
symmetry breaking. But the $2N$ levels split into bands when quantum
fluctuations arise with $h>0$, which prevents the spontaneous symmetry
breaking and thus protects the odd parity of the ground state.

\emph{Finite temperature properties of the gapless phase.}---In the gapless
phase, the lowest $2N$ neatly aligned levels dominate the system's properties
at low temperatures ($T$ $\ll4h/k_{B}$), where $k_{B}$ is the Boltzmann
constant. This fact facilitates us to work out some quantities at low
temperatures based on the partition function $Z=\sum_{q\in q^{o}%
\;\text{and}\;q^{e}}\operatorname{e}^{-\beta E_{q}}$, where $\beta=\frac
{1}{k_{B}T}$. The density of states (DOS) is defined as $\rho(x)=\frac{1}%
{N}\sum_{q\in q^{o}\;\text{and}\;q^{e}}\,\delta(E-E_{q})$. Considering the
thermodynamic limit, we get a DOS at low energy sector
\begin{equation}
\rho(x)=\frac{4\left(  x+2J-2h\right)  }{\pi\sqrt{x\left(  x-4h\right)
\left(  4h-x-4J\right)  \left(  x+4J\right)  }},
\end{equation}
which can be expanded as $\rho(x)$=$ax^{-1/2}+bx^{1/2}+O(x^{3/2})$ with
$x=E-E_{0}$, $a$=$\frac{(J-h)^{1/2}}{\pi(Jh)^{1/2}}$, and $b$=$\frac
{(h^{2}+Jh+J^{2})}{8\pi(Jh)^{3/2}(J-h)^{1/2}}$. So we get the specific heat
per site at low temperatures,
\begin{equation}
\frac{C_{M}(T)}{N}=\frac{k_{B}}{2}\left[  1+\frac{2bk_{B}T(4a+bk_{B}%
T)}{(2a+bk_{B}T)^{2}}\right]  , \label{CM}%
\end{equation}
which reaches a constant $k_{B}/2$ at zero temperature. Next, we consider the
correlation function at low temperatures. When we evaluated the correlation
function for each of the lowest $2N$ levels, we got an unexpected finding:
$\langle E_{q}|\sigma_{j}^{x}\sigma_{j+r}^{x}|E_{q}\rangle$=$\langle
E_{0}|\sigma_{j}^{x}\sigma_{j+r}^{x}|E_{0}\rangle$ for $q\in q^{o}%
\;$and$\;q^{e}$, although the resulting Toeplitz determinants are a little
different \cite{S1}. So we draw a conclusion that the correlation function for
large enough $r$ and $N$ is inert to the change of temperature at low
temperatures,%
\begin{equation}
C_{r,N}^{xx}(0\leq T\ll4h/k_{B})=C_{r,N}^{xx}.
\end{equation}

\emph{Experimental proposal.}---Although it is not possible to get an infinite
FTIR system to observe its fascinating properties, we may design a large
enough one to see the trend and realize the fascinating states within nowadays
state-of-art techniques based on laser-cooled and trapped atomic ions. In
fact, the case for $N=3$ has been experimentally realized \cite{Edwards}. To
generate a system with larger $N$, we provide another proposal.

\begin{figure}[ptb]
\includegraphics[width=0.4\textwidth]{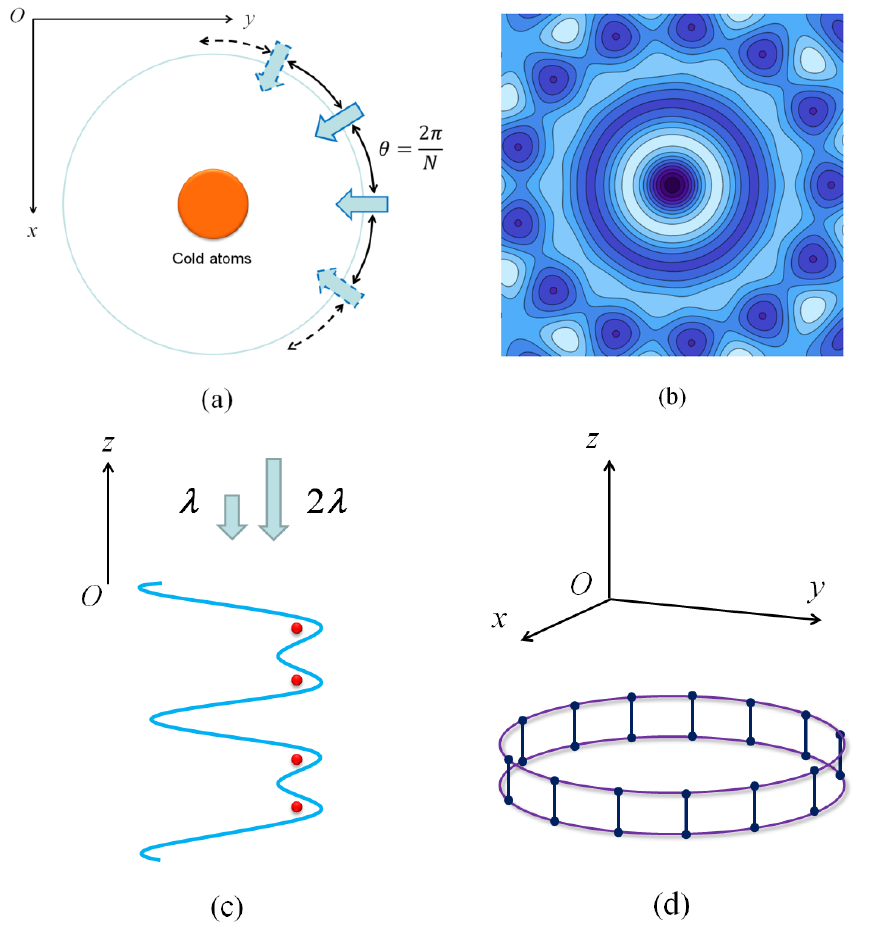}\newline\caption{(Color online)
(a) Scheme of the proposed experimental setup in $x$-$y$ plane. Each
arrow depicts a wave vector of a standing-wave laser. The angle between any
two neighboring lasers is $2\pi/N$. (b) The exemplified color map of optical
potential where a ring of $13$ trapping wells is shown by the dark blue
potential wells. (c) The arrangement of lasers in $z$ direction, where two
standing wave lasers form an isolated double wells potential. (d) The total
two-leg ladder potential.}
\label{fig5}
\end{figure}

In our proposal as shown in Fig. \ref{fig5}, the key point is to produce a
ring geometrical potential with odd number of traps. In $x$-$y$ plane, we need
to impose $N$ (odd) beams of independent standing wave lasers which are
obtained by frequency selection. Then, each standing wave will contributes an
optical potential along $\overrightarrow{k_{i}}$\ direction that can be
expressed as $V_{x\text{-}y}\cos^{2}(\overrightarrow{k_{i}}\cdot
\overrightarrow{r_{i}}-\phi)$\ for the i-th beam, where $\overrightarrow{k_{i}%
}$ is the strength of beams and $\phi$ is the phase shift. The angle between
two neighboring lasers is $2\pi/N$. Thus, by adopting $V_{x\text{-}y}$\ and
$\phi$, we obtain a circular lattice potential with $N$ traps in $x$-$y$
plane. In $z$ direction we apply two independent standing wave lasers,
$V_{z1}\cos^{2}(k_{z}z)$ and $V_{z2}\cos^{2}(2k_{z}z)$, where the former has
twice wave length of the latter. Eventually, we obtain a periodical two-leg
ladder potential by forming a double-well potential in $z$ direction.

Next, we consider loading into the ladders with cold atoms having two relevant
internal states. For sufficiently deep potential and low temperatures, the
system will be described by a bosonic or fermionic Hubbard model \cite{Duan}.
In the Mott-insulating phase, it can be reduced to a pseudo-spin $XXZ$ model
by second-order perturbation at half-filling. By modulating the intensity and
the phase shift of the trapping laser beams and by adjusting scattering length
through Feshbach resonance, we can eventually arrive at the pseudo-spin
Hamiltonian \cite{S1},%
\begin{equation}
H_{s}=\sum_{j,s=1,2}J_{z}S_{j,s}^{z}S_{j+1,s}^{z}+\sum_{j}K_{{}}\vec{S}%
_{j,1}\cdot\vec{S}_{j,2},
\end{equation}
The low-energy properties of this system are dominated by the pseudo-spin
singlet $\left\vert s\right\rangle _{j}$=$(\left\vert \uparrow\downarrow
\right\rangle _{j}-\left\vert \downarrow\uparrow\right\rangle _{j})/\sqrt{2}$
and triplet $\left\vert t_{0}\right\rangle _{j}$=$(\left\vert \uparrow
\downarrow\right\rangle _{j}+\left\vert \downarrow\uparrow\right\rangle
_{j})\sqrt{2}$\ on the rung of the ladders in low energy. At this time, the
system can be mapped exactly to an effective FTIR \cite{ChenQH}.

\emph{Conclusion.}---In summary, we have shown an unusual gapless phase
induced by the ring frustration in the FTIR. We demonstrated a novel behavior
of the correlation function in the gapless phase at both zero and low
temperatures. The ground state in the gapless phase is
dominated by a superposition of all possible antiferromagnetic kink states,
which results in a quite large entropy of entanglement.
The non-degeneracy of the ground state prevents spontaneous symmetry breaking.
These unexpected outcomes are in close relation to the odevity of the number
of lattice sites even in thermodynamic limit. If given an even number of
lattice sites, the gapless phase would turn into a gapped one as the usual
schematic way \cite{Sachdev}. The odevity-induced phenomenon revealed in this
Letter is reminiscent of the one in the well-known spin ladders \cite{Dagotto}%
. We also proposed an experimental protocol for observing the peculiar gapless phase.

We acknowledge useful discussions with Yan He. This work was supported by the
NSFC under Grants no. 11074177, SRF for ROCS SEM (20111139-10-2).

\clearpage

\onecolumngrid
\begin{appendix}

\renewcommand\thesection{\arabic{section}}
\renewcommand\thefigure{S\arabic{figure}}
\renewcommand\theequation{S\arabic{equation}}
\setcounter{figure}{0}
\setcounter{equation}{0}

\section{Supplemental material}

\section{I. \quad The correlation function in the gapless phase}

\label{sec:cor}

\subsection{Proposition and proof}

\noindent\textit{Proposition: }Consider a Toeplitz determinant%
\begin{equation}
\Theta(r,N,x,\operatorname{e}^{\operatorname*{i}k})=\left\vert
\begin{array}
[c]{cccc}%
\overset{\sim}{D}_{0} & \overset{\sim}{D}_{-1} & \cdots & \overset{\sim
}{D}_{1-r}\\
\overset{\sim}{D}_{1} & \overset{\sim}{D}_{0} & \cdots & \overset{\sim
}{D}_{2-r}\\
\cdots & \cdots & \cdots & \cdots\\
\overset{\sim}{D}_{r-1} & \overset{\sim}{D}_{r-2} & \cdots & \overset{\sim
}{D}_{0}%
\end{array}
\right\vert \label{sTheta}%
\end{equation}
with $\overset{\sim}{D}_{n}$=$\int_{-\pi}^{\pi}\frac{dq}{2\pi}%
\,D(\operatorname{e}^{\operatorname*{i}q})\,\operatorname{e}%
^{-\operatorname*{i}qn}+\frac{x}{N}\operatorname{e}^{\operatorname*{i}kn} $.
If the generating function $D(\operatorname{e}^{\operatorname*{i}q})$ and $\ln
D(\operatorname{e}^{\operatorname*{i}q})$\ are continuous on the unit circle
$\left\vert \operatorname{e}^{\operatorname*{i}q}\right\vert =1$, then the
behavior for large $N$ of $\Theta(r,N,x,\operatorname{e}^{\operatorname*{i}%
k})$ is given by ($1\ll r\ll N$)
\begin{equation}
\Theta(r,N,x,\operatorname{e}^{\operatorname*{i}k})=\Delta_{r}(1+\frac
{xr}{ND(\operatorname{e}^{-\operatorname*{i}k})}),
\end{equation}
where
\begin{align}
\Delta_{r}  &  =\mu^{r}\exp(\sum_{n=1}^{\infty}nd_{-n}d_{n}),\label{sdeltar}\\
\mu &  =\exp\left[  \int_{-\pi}^{\pi}\frac{dq}{2\pi}\,\ln D(\operatorname{e}%
^{\operatorname*{i}q})\right]  ,\\
d_{n}  &  =\int_{-\pi}^{\pi}\frac{dq}{2\pi}\,e^{-\operatorname*{i}qn}\ln
D(\operatorname{e}^{\operatorname*{i}q}) ,
\end{align}
if the sum in Eq. (\ref{sdeltar}) converges.\newline\strut\noindent
\textit{Proof: }Let\textit{\ }$\operatorname{e}^{\operatorname*{i}q}=\xi,$
$D_{n}=\int_{-\pi}^{\pi}\frac{dq}{2\pi}\,D(\xi)\xi^{-n}\,,$ then
$\overset{\sim}{D}_{n}=D_{n}+\frac{x}{N}\operatorname{e}^{\operatorname*{i}%
kn}.$ First, we rewrite Eq. (\ref{sTheta}) as%
\begin{align*}
\Theta(r,N,x,\operatorname{e}^{\operatorname*{i}k})  &  =\left\vert
\begin{array}
[c]{cccc}%
D_{0} & D_{-1} & \cdots & D_{-r+1}\\
D_{1} & D_{0} & \cdots & D_{-r+2}\\
\cdots & \cdots & \cdots & \cdots\\
D_{r-1} & D_{r-2} & \cdots & D_{0}%
\end{array}
\right\vert +\left\vert
\begin{array}
[c]{cccc}%
\frac{x}{N} & D_{-1} & \cdots & D_{1-r}\\
\frac{x}{N}\operatorname{e}^{\operatorname*{i}k} & D_{0} & \cdots & D_{2-r}\\
\cdots & \cdots & \cdots & \cdots\\
\frac{x}{N}\operatorname{e}^{\operatorname*{i}(r-1)k} & D_{r-2} & \cdots &
D_{0}%
\end{array}
\right\vert \\
&  +\ldots+\left\vert
\begin{array}
[c]{cccc}%
D_{0} & \frac{x}{N}\operatorname{e}^{-\operatorname*{i}k} & \cdots & D_{2-r}\\
D_{1} & \frac{x}{N} & \cdots & D_{2-r}\\
\cdots & \cdots & \cdots & \cdots\\
D_{r-1} & \frac{x}{N}\operatorname{e}^{\operatorname*{i}(r-2)k} & \cdots &
D_{0}%
\end{array}
\right\vert +\left\vert
\begin{array}
[c]{cccc}%
D_{0} & D_{-1} & \cdots & \frac{x}{N}\operatorname{e}^{\operatorname*{i}%
(1-r)k}\\
D_{1} & D_{0} & \cdots & \frac{x}{N}\operatorname{e}^{\operatorname*{i}%
(2-r)k}\\
\cdots & \cdots & \cdots & \cdots\\
D_{r-1} & D_{r-2} & \cdots & \frac{x}{N}%
\end{array}
\right\vert .
\end{align*}
Then we compose a set of linear equations
\begin{equation}
\sum_{m=0}^{r-1}D_{n-m}x_{m}^{(r-1)}=\frac{x}{N}\mathrm{e}^{\mathrm{i}%
kn}\;,\;0\leq n\leq r-1 .\label{sumeq}%
\end{equation}
These equations have an unique solution for $x_{n}^{(r-1)}$ if there exists a
non-zero determinant:
\begin{equation}
\Delta_{r}\equiv\left\vert
\begin{array}
[c]{cccc}%
D_{0} & D_{-1} & \cdots & D_{1-r}\\
D_{1} & D_{0} & \cdots & D_{2-r}\\
\cdots & \cdots & \cdots & \cdots\\
D_{r-1} & D_{r-2} & \cdots & D_{0}%
\end{array}
\right\vert \neq0 .
\end{equation}
By Cramer's rule, we have the solution:
\begin{align}
&  x_{0}^{(r-1)}=\frac{\left\vert
\begin{array}
[c]{cccc}%
\frac{x}{N} & D_{-1} & \cdots & D_{1-r}\\
\frac{x}{N}\operatorname{e}^{\operatorname*{i}k} & D_{0} & \cdots & D_{2-r}\\
\cdots & \cdots & \cdots & \cdots\\
\frac{x}{N}\operatorname{e}^{\operatorname*{i}(r-1)k} & D_{r-2} & \cdots &
D_{0}%
\end{array}
\right\vert }{\Delta_{r}}\;,\\
&  x_{1}^{(r-1)}=\frac{\left\vert
\begin{array}
[c]{cccc}%
D_{0} & \frac{x}{N}\operatorname{e}^{-\operatorname*{i}k} & \cdots & D_{2-r}\\
D_{1} & \frac{x}{N} & \cdots & D_{2-r}\\
\cdots & \cdots & \cdots & \cdots\\
D_{r-1} & \frac{x}{N}\operatorname{e}^{\operatorname*{i}(r-2)k} & \cdots &
D_{0}%
\end{array}
\right\vert }{\Delta_{r}}\;,\\
&  \qquad\qquad\qquad\qquad\qquad\vdots\nonumber\\
&  x_{r-1}^{(r-1)}=\frac{\left\vert
\begin{array}
[c]{cccc}%
D_{0} & D_{-1} & \cdots & \frac{x}{N}\operatorname{e}^{\operatorname*{i}%
(1-r)k}\\
D_{1} & D_{0} & \cdots & \frac{x}{N}\operatorname{e}^{\operatorname*{i}%
(2-r)k}\\
\cdots & \cdots & \cdots & \cdots\\
D_{r-1} & D_{r-2} & \cdots & \frac{x}{N}%
\end{array}
\right\vert }{\Delta_{r}} .
\end{align}
So we arrive at
\begin{equation}
\Theta(r,N,x,\operatorname{e}^{\operatorname*{i}k})=\Delta_{r}+\Delta_{r}%
\sum_{n=0}^{r-1}\operatorname{e}^{-\operatorname*{i}kn}x_{n}^{(r-1)}
.\label{sumxn}%
\end{equation}
For our problem, $\Delta_{r}$ can be evaluated directly by using Szeg\"{o}'s
theorem, so we need to know how to calculate the second term in Eq.
(\ref{sumxn}). Follow the standard Wiener-Hopf procedure \cite{McCoy Wu,McCoy,
Wu}, we consider a generalization of Eq. (\ref{sumeq})
\begin{equation}
\sum_{m=0}^{r-1}D_{n-m}x_{m}=y_{n}\text{, \ }0\leq n\leq r-1\label{Wiener}%
\end{equation}
and define
\begin{equation}
x_{n}=y_{n}=0\quad\text{for}\quad n\leq-1\quad\text{and}\quad n\geq r
\end{equation}%
\begin{align}
v_{n}  &  =\sum_{m=0}^{r-1}D_{-n-m}x_{m}\quad\text{for}\quad n\geq1\nonumber\\
&  =0\quad\text{for}\quad n\leq0
\end{align}%
\begin{align}
u_{n}  &  =\sum_{m=0}^{r-1}D_{r-1+n-m}x_{m}\quad\text{for}\quad n\geq
1\nonumber\\
&  =0\quad\text{for}\quad n\leq0
\end{align}
We further define
\begin{equation}
D\left(  \xi\right)  =\sum_{n=-\infty}^{\infty}D_{n}\xi^{n},\quad Y\left(
\xi\right)  =\sum_{n=0}^{r-1}y_{n}\xi^{n},\quad V\left(  \xi\right)
=\sum_{n=1}^{\infty}v_{n}\xi^{n},\quad U\left(  \xi\right)  =\sum
_{n=1}^{\infty}u_{n}\xi^{n},\quad X\left(  \xi\right)  =\sum_{n=0}^{r-1}%
x_{n}\xi^{n} .\label{fourier}%
\end{equation}
It then follows from Eq. (\ref{Wiener}) that we can get
\begin{equation}
D\left(  \xi\right)  X\left(  \xi\right)  =Y\left(  \xi\right)  +V\left(
\xi^{-1}\right)  +U\left(  \xi\right)  \xi^{r-1}\label{fourier form}%
\end{equation}
for $|\xi|=1$. Becuase $D\left(  \xi\right)  $ and $\ln D\left(  \xi\right)  $
is continuous and periodic on the unit circle, $D\left(  \xi\right)  $ has a
unique factorization, up to a multiplicative constant, in the form
\begin{equation}
D\left(  \xi\right)  =P^{-1}\left(  \xi\right)  Q^{-1}\left(  \xi^{-1}\right)
,\label{Dxi}%
\end{equation}
for $|\xi|=1$, such that $P\left(  \xi\right)  $ and $Q\left(  \xi\right)  $
are both analytic for $|\xi|<1$ and continuous and nonzero for $|\xi|\leq1$.
we may now use the factorization of $D\left(  \xi\right)  $ in Eq.
(\ref{fourier form}) to write
\begin{align}
&  P^{-1}\left(  \xi\right)  X\left(  \xi\right)  -\left[  Q\left(  \xi
^{-1}\right)  Y\left(  \xi\right)  \right]  _{+}-\left[  Q\left(  \xi
^{-1}\right)  U\left(  \xi\right)  \xi^{r-1}\right]  _{+}\nonumber\\
&  =\left[  Q\left(  \xi^{-1}\right)  Y\left(  \xi\right)  \right]
_{-}+Q\left(  \xi^{-1}\right)  V\left(  \xi^{-1}\right)  +\left[  Q\left(
\xi^{-1}\right)  U\left(  \xi\right)  \xi^{r-1}\right]  _{-} ,\label{pq}%
\end{align}
where the subscript $+\left(  -\right)  $ means that we should expand the
quantity in the brackets into a Laurent series and keep only those terms where
$\xi$ is raised to a non-negative (negative) power. The left-hand side of Eq.
(\ref{pq}) defines a function analytic for $|\xi|<1$ and continuous on
$|\xi|=1$ and the right-hand side defines a function which is analytic for
$|\xi|>1$ and is continuous for $|\xi|=1$. Taken together they define a
function $E(\xi)$ analytic for all $\xi$ except possibly for $|\xi|=1$ and
continuous everywhere. But these properties are sufficient to prove that
$E(\xi)$ is an entire function which vanished at $|\xi|=\infty$ and thus, by
Liouville's theorem, must be zero everywhere \cite{McCoy Wu,McCoy}. Therefore
both the right-hand side and the left-hand side of Eq. (\ref{pq}) vanish
separately and thus we have
\begin{equation}
X\left(  \xi\right)  =P\left(  \xi\right)  \left\{  \left[  Q\left(  \xi
^{-1}\right)  Y\left(  \xi\right)  \right]  _{+}+\left[  Q\left(  \xi
^{-1}\right)  U\left(  \xi\right)  \xi^{r-1}\right]  _{+}\right\} .
\end{equation}
Furthermore, $U\left(  \xi\right)  $ can be neglected for large r%
\begin{equation}
X\left(  \xi\right)  \approx P\left(  \xi\right)  \left[  Q\left(  \xi
^{-1}\right)  Y\left(  \xi\right)  \right]  _{+} .\label{xxi}%
\end{equation}
Consider the term $\left[  Q\left(  \xi^{-1}\right)  Y\left(  \xi\right)
\right]  _{+}$, because $Q\left(  \xi\right)  $ is a $+$ function, so we can
expand it as a Laurent series and keep only those term where $\xi$ is raised
to a non-negative power,
\begin{equation}
Q\left(  \xi\right)  = {\displaystyle\sum\limits_{n=0}^{\infty}} a_{n}\xi
^{n}=\left(  a_{0}+a_{1}\xi^{1}+a_{2}\xi^{2}+\cdots+a_{r-1}\xi^{r-1}\right)
+O\left(  \xi^{r}\right) ,
\end{equation}
and then
\begin{equation}
Q\left(  \xi^{-1}\right)  =a_{0}+a_{1}\xi^{-1}+a_{2}\xi^{-2}+\cdots+a_{r-1}%
\xi^{1-r},
\end{equation}
where we have neglected the term $O\left(  \xi^{r}\right)  $\ for large $r$
for clarity, from Eq. (\ref{sumeq}) and Eq. (\ref{fourier}), we have
\begin{equation}
Y\left(  \xi\right)  =\sum_{n=0}^{r-1}y_{n}\xi^{n}=\frac{x}{N}\left(
1+\operatorname{e}^{\operatorname*{i}k}\xi^{1}+\operatorname{e}%
^{2\operatorname*{i}k}\xi^{2}+\cdots+\operatorname{e}^{\operatorname*{i}%
\left(  r-1\right)  k}\xi^{r-1}\right) ,
\end{equation}
thus
\begin{align}
\left[  Q\left(  \xi^{-1}\right)  Y\left(  \xi\right)  \right]  _{+}  &
=\frac{x}{N}[\left(  a_{0}+a_{1}\operatorname{e}^{\operatorname*{i}k}%
+a_{2}\operatorname{e}^{2\operatorname*{i}k}+\cdots+a_{r-1}\operatorname{e}%
^{\operatorname*{i}\left(  r-1\right)  k}\right) \nonumber\\
&  \mathstrut\quad+\left(  a_{0}\operatorname{e}^{\operatorname*{i}k}%
+a_{1}\operatorname{e}^{2\operatorname*{i}k}+\cdots+a_{r-2}\operatorname{e}%
^{\operatorname*{i}\left(  r-1\right)  k}\right)  \xi^{1}+\cdots+\left(
a_{0}\operatorname{e}^{\operatorname*{i}\left(  r-1\right)  k}\right)
\xi^{r-1}].
\end{align}
From Eq. (\ref{sumxn}), Eq. (\ref{fourier}) and Eq. (\ref{xxi}), we have%
\begin{equation}
\sum_{n=0}^{r-1}\operatorname{e}^{-\operatorname*{i}kn}x_{n}^{(r-1)}=X\left(
\operatorname{e}^{-\operatorname*{i}k}\right)  =P\left(  \operatorname{e}%
^{-\operatorname*{i}k}\right)  \left[  Q\left(  \operatorname{e}%
^{\operatorname*{i}k}\right)  Y\left(  \operatorname{e}^{-\operatorname*{i}%
k}\right)  \right]  _{+},
\end{equation}
\begin{align}
\left[  Q\left(  \operatorname{e}^{\operatorname*{i}k}\right)  Y\left(
\operatorname{e}^{-\operatorname*{i}k}\right)  \right]  _{+}  &  =\frac{x}%
{N}\left[  ra_{0}+ra_{1}\operatorname{e}^{\operatorname*{i}k}+ra_{2}%
\operatorname{e}^{2\operatorname*{i}k}+\cdots+ra_{r-1}\operatorname{e}%
^{\operatorname*{i}\left(  r-1\right)  k}\right] \nonumber\\
&  \mathstrut\quad-\frac{x}{N}\left[  a_{1}\operatorname{e}^{\operatorname*{i}%
k}+2a_{2}\operatorname{e}^{2\operatorname*{i}k}+\cdots+\left(  r-1\right)
a_{r-1}\operatorname{e}^{\operatorname*{i}\left(  r-1\right)  k}\right]
\nonumber\\
&  =\frac{x}{N}\left[  rQ\left(  \operatorname{e}^{\operatorname*{i}k}\right)
-\operatorname{e}^{\operatorname*{i}k}\frac{dQ\left(  \xi\right)  }{d\xi
}|_{\xi=\operatorname{e}^{\operatorname*{i}k}}\right]  .\label{qy}%
\end{align}
So when $r\gg1$, we can ignore the second term in Eq. (\ref{qy}). Together
with Eq. (\ref{Dxi}), we get
\begin{equation}
X\left(  \operatorname{e}^{-\operatorname*{i}k}\right)  =\frac{xr}{N}P\left(
\operatorname{e}^{-\operatorname*{i}k}\right)  Q\left(  \operatorname{e}%
^{\operatorname*{i}k}\right)  =\frac{xr}{ND\left(  \operatorname{e}%
^{-\operatorname*{i}k}\right)  }.
\end{equation}
At last, by Szeg\"{o}'s theorem, we get
\begin{equation}
\Delta_{r}=\mu^{r}\exp(\sum_{n=1}^{\infty}nd_{-n}d_{n}),
\end{equation}
where%
\begin{align}
\mu &  =\exp\left[  \int_{-\pi}^{\pi}\frac{dq}{2\pi}\,\ln D(\operatorname{e}%
^{\operatorname*{i}q})\right]  ,\nonumber\\
d_{n}  &  =\int_{-\pi}^{\pi}\frac{dq}{2\pi}\,e^{-\operatorname*{i}qn}\ln
D(\operatorname{e}^{\operatorname*{i}q}).
\end{align}
From Eq. (\ref{sumxn}), we have%
\begin{equation}
\Theta(r,N,x,\operatorname{e}^{\operatorname*{i}k})=\Delta_{r}\left(
1+\frac{xr}{ND\left(  \operatorname{e}^{-\operatorname*{i}k}\right)  }\right)
.
\end{equation}
Q.E.D.

\subsection{Application to the FTIR: the correlation functions}

We now specialize to the problem of the FTIR in the main text. We calculate
the correlation functions of the lowest $2N$ levels that are grouped into four
bands. The first band with indexes $\left(  -1,1,1,0\right)  $ is the ground
state coming from the odd channel, $|E_{0}\rangle=c_{0}^{\dag}|\phi^{o}%
\rangle$. We have
\begin{align}
C_{r,N}^{x,x}\left(  |E_{0}\rangle\right)   &  =\langle\phi^{o}|c_{0}%
B_{j}A_{j+1}B_{j+1}\ldots B_{j+r-1}A_{j+r}c_{0}^{\dag}|\phi^{o}\rangle
\nonumber\\
&  =\left\vert
\begin{array}
[c]{cccc}%
D_{0}+\frac{2}{N} & D_{-1}+\frac{2}{N} & \cdots & D_{1-r}+\frac{2}{N}\\
D_{1}+\frac{2}{N} & D_{0}+\frac{2}{N} & \cdots & D_{2-r}+\frac{2}{N}\\
\cdots & \cdots & \cdots & \cdots\\
D_{r-1}+\frac{2}{N} & D_{r-2}+\frac{2}{N} & \cdots & D_{0}+\frac{2}{N}%
\end{array}
\right\vert ,\label{corgs}%
\end{align}
where%
\begin{equation}
D_{n}=\frac{1}{N}\sum_{_{\substack{q\in q^{o},q\neq0}}}\exp\left(
\operatorname*{i}q\left(  n-1\right)  \right)  \left(  1-2u_{q}^{2}%
-2\operatorname*{i}u_{q}v_{q}\right)  -\frac{1}{N}.\label{Dnodd}%
\end{equation}
The second band with indexes $\left(  1,2,0,1\right)  $ is a collection of
states from the even channel, $|E_{k}\rangle=\eta_{k}^{\dagger}c_{\pi
}^{\dagger}|\phi^{e}\rangle$ ($k\in q^{e}$, $k\neq\pi$). We have%
\begin{align}
C_{r,N}^{x,x}\left(  |E_{k}\rangle\right)   &  =\frac{1}{2}\left(  \langle
\phi^{e}|c_{\pi}\eta_{k}B_{j}A_{j+1}\ldots B_{j+r-1}A_{j+r}\eta_{k}^{\dagger
}c_{\pi}^{\dagger}|\phi^{e}\rangle+\langle\phi^{e}|c_{\pi}\eta_{-k}%
B_{j}A_{j+1}\ldots B_{j+r-1}A_{j+r}\eta_{-k}^{\dagger}c_{\pi}^{\dagger}%
|\phi^{e}\rangle\right) \nonumber\\
&  =\frac{1}{2}\left[  \Gamma^{e}(r,N,\alpha_{k},\operatorname{e}%
^{\operatorname*{i}k})+\Gamma^{e}(r,N,\alpha_{-k},\operatorname{e}%
^{-\operatorname*{i}k})\right]  ,\label{corqeven}%
\end{align}
where%
\begin{equation}
\Gamma^{e}(r,N,\alpha_{k},\operatorname{e}^{\operatorname*{i}k})=\left\vert
\begin{array}
[c]{cccc}%
F_{0}+\frac{2\alpha^{k}}{N} & F_{-1}+\frac{2\alpha^{k}}{N}\operatorname{e}%
^{-\operatorname*{i}k} & \cdots & F_{1-r}+\frac{2\alpha^{k}}{N}%
\operatorname{e}^{\operatorname*{i}(1-r)k}\\
F_{1}+\frac{2\alpha^{k}}{N}\operatorname{e}^{\operatorname*{i}k} & F_{0}%
+\frac{2\alpha^{k}}{N} & \cdots & F_{2-r}+\frac{2\alpha^{k}}{N}%
\operatorname{e}^{\operatorname*{i}(2-r)k}\\
\cdots & \cdots & \cdots & \cdots\\
F_{r-1}+\frac{2\alpha^{k}}{N}\operatorname{e}^{\operatorname*{i}(r-1)k} &
F_{r-2}+\frac{2\alpha^{k}}{N}\operatorname{e}^{\operatorname*{i}(r-2)k} &
\cdots & F_{0}+\frac{2\alpha^{k}}{N}%
\end{array}
\right\vert ,\label{Gammae}%
\end{equation}%
\begin{equation}
\langle\phi^{e}|B_{l}A_{m}|\phi^{e}\rangle=F_{l-m+1}=\frac{1}{N}\sum_{
_{\substack{q\in q^{e}}}}\exp(iq\left(  l-m\right)  )(1-2u_{q}^{2}%
-2iu_{q}v_{q}).\label{Feven}%
\end{equation}%
\begin{equation}
\alpha_{k}=\frac{J-h\operatorname{e}^{-\operatorname*{i}k}}{\sqrt{\left(
J-h\operatorname{e}^{-\operatorname*{i}k}\right)  \left(  J-h\operatorname{e}%
^{\operatorname*{i}k}\right)  }}.\label{alphakodd}%
\end{equation}
Notice that we have defined $u_{\pi}^{2}=0,2u_{\pi}v_{\pi}=0$ in Eq.
(\ref{Feven}). The third band with indexes $\left(  -1,1,0,0\right)  $ is a
collection of states from the odd channel, $|E_{k}\rangle=\eta_{k}^{\dagger
}|\phi^{o}\rangle$($k\in q^{o}$, $k\neq0$). We have%
\begin{align}
C_{r,N}^{x,x}\left(  |E_{k}\rangle\right)   &  =\frac{1}{2}\left(  \langle
\phi^{o}|\eta_{k}B_{j}A_{j+1}\ldots B_{j+r-1}A_{j+r}\eta_{k}^{\dagger}%
|\phi^{o}\rangle+\langle\phi^{o}|\eta_{-k}B_{j}A_{j+1}\ldots B_{j+r-1}%
A_{j+r}\eta_{-k}^{\dagger}|\phi^{o}\rangle\right) \nonumber\\
&  =\frac{1}{2}\left[  \Gamma^{o}(r,N,\alpha_{k},\operatorname{e}%
^{\operatorname*{i}k})+\Gamma^{o}(r,N,\alpha_{-k},\operatorname{e}%
^{-\operatorname*{i}k})\right]  ,\label{corqodd}%
\end{align}
where%
\begin{equation}
\Gamma^{o}(r,N,\alpha_{k},\operatorname{e}^{\operatorname*{i}k})=\left\vert
\begin{array}
[c]{cccc}%
D_{0}+\frac{2\alpha^{k}}{N} & D_{-1}+\frac{2\alpha^{k}}{N}\operatorname{e}%
^{-\operatorname*{i}k} & \cdots & D_{1-r}+\frac{2\alpha^{k}}{N}%
\operatorname{e}^{\operatorname*{i}(1-r)k}\\
D_{1}+\frac{2\alpha^{k}}{N}\operatorname{e}^{\operatorname*{i}k} & D_{0}%
+\frac{2\alpha^{k}}{N} & \cdots & D_{2-r}+\frac{2\alpha^{k}}{N}%
\operatorname{e}^{\operatorname*{i}(2-r)k}\\
\cdots & \cdots & \cdots & \cdots\\
D_{r-1}+\frac{2\alpha^{k}}{N}\operatorname{e}^{\operatorname*{i}(r-1)k} &
D_{r-2}+\frac{2\alpha^{k}}{N}\operatorname{e}^{\operatorname*{i}(r-2)k} &
\cdots & D_{0}+\frac{2\alpha^{k}}{N}%
\end{array}
\right\vert ,\label{Gamma}%
\end{equation}
with $\alpha_{k}$ $\left(  k\in q^{o},k\neq0\right)  $ given by Eq.
(\ref{alphakodd}). The last band with indexes $\left(  1,0,0,0\right)  $ is
the state from the even channel, $|E_{\pi}\rangle=|\phi^{e}\rangle$. We have
\begin{align}
C_{r,N}^{x,x}\left(  |E_{\pi}\rangle\right)   &  =\langle\phi^{e}|B_{j}%
A_{j+1}B_{j+1}\ldots A_{j+r-1}B_{j+r-1}A_{j+r}|\phi^{e}\rangle\nonumber\\
&  =\left\vert
\begin{array}
[c]{cccc}%
F_{0}+\frac{2}{N} & F_{-1}+\frac{2}{N}\operatorname{e}^{-\operatorname*{i}\pi}
& \cdots & F_{1-r}+\frac{2}{N}\operatorname{e}^{\operatorname*{i}\pi\left(
1-r\right)  }\\
F_{1}+\frac{2}{N}\operatorname{e}^{\operatorname*{i}\pi} & F_{0}+\frac{2}{N} &
\cdots & F_{2-r}+\frac{2}{N}\operatorname{e}^{\operatorname*{i}\pi\left(
2-r\right)  }\\
\cdots & \cdots & \cdots & \cdots\\
F_{r-1}+\frac{2}{N}\operatorname{e}^{\operatorname*{i}\pi\left(  r-1\right)  }
& F_{r-2}+\frac{2}{N}\operatorname{e}^{\operatorname*{i}\pi\left(  r-2\right)
} & \cdots & F_{0}+\frac{2}{N}%
\end{array}
\right\vert ,\label{corpi}%
\end{align}
Note that Eq. (\ref{corgs}), Eq. (\ref{corqeven}), Eq. (\ref{corqodd}) and Eq.
(\ref{corpi}) are valid for the gapped phase, the critical point and the
gapless phase. For the most interesting gapless phase, we have $u_{0}%
^{2}=1,2u_{0}v_{0}=0$. we can rewrite Eq. (\ref{Dnodd}) as%
\begin{equation}
D_{n}=\frac{1}{N}\sum_{_{\substack{q\in q^{o}}}}\exp\left(  \operatorname*{i}%
q\left(  n-1\right)  \right)  \left(  1-2u_{q}^{2}-2\operatorname*{i}%
u_{q}v_{q}\right)  .
\end{equation}
For large $N$, we have
\begin{equation}
D_{n}=F_{n}=\int_{-\pi}^{\pi}\frac{dq}{2\pi}\operatorname{e}%
^{-\operatorname*{i}qn}\frac{-\left(  J-h\operatorname{e}^{\operatorname*{i}%
q}\right)  }{\sqrt{\left(  J-h\operatorname{e}^{-\operatorname*{i}q}\right)
\left(  J-h\operatorname{e}^{\operatorname*{i}q}\right)  }}.
\end{equation}
So the above four cases of correlation function in the gapless phase can be
written in an uniform formula%
\begin{equation}
C_{r,N}^{x,x}\left(  |E_{q}\rangle\right)  =\frac{1}{2}\left[  \Gamma
(r,N,\alpha_{k},\operatorname{e}^{\operatorname*{i}k})+\Gamma(r,N,\alpha
_{-k},\operatorname{e}^{-\operatorname*{i}k})\right]  ,
\end{equation}
where
\begin{equation}
\Gamma(r,N,\alpha_{k},\operatorname{e}^{\operatorname*{i}k})=\left\vert
\begin{array}
[c]{cccc}%
\overset{\sim}{D}_{0} & \overset{\sim}{D}_{-1} & \cdots & \overset{\sim
}{D}_{1-r}\\
\overset{\sim}{D}_{1} & \overset{\sim}{D}_{0} & \cdots & \overset{\sim
}{D}_{2-r}\\
\cdots & \cdots & \cdots & \cdots\\
\overset{\sim}{D}_{r-1} & \overset{\sim}{D}_{r-2} & \cdots & \overset{\sim
}{D}_{0}%
\end{array}
\right\vert ,
\end{equation}%
\[
\overset{\sim}{D}_{n}=\int_{-\pi}^{\pi}\frac{dq}{2\pi}\,\operatorname{e}%
^{-\operatorname*{i}qn}\frac{-\left(  J-h\operatorname{e}^{\operatorname*{i}%
q}\right)  }{\sqrt{\left(  J-h\operatorname{e}^{-\operatorname*{i}q}\right)
\left(  J-h\operatorname{e}^{\operatorname*{i}q}\right)  }}+\frac{2\alpha^{k}%
}{N}\operatorname{e}^{\operatorname*{i}kn},
\]%
\begin{equation}
\alpha^{k}=\frac{J-h\operatorname{e}^{-\operatorname*{i}k}}{\sqrt{\left(
J-h\operatorname{e}^{-\operatorname*{i}k}\right)  \left(  J-h\operatorname{e}%
^{\operatorname*{i}k}\right)  }}\;,\quad\text{with}\quad k\in\left\{
\begin{array}
[c]{ll}%
k^{o}=-\frac{(N-1)\pi}{N},\ldots,-\frac{2\pi}{N},0,\frac{2\pi}{N},\ldots
,\frac{(N-1)\pi}{N} & \text{for odd channel}\\
k^{e}=-\frac{(N-2)\pi}{N},\ldots,-\frac{\pi}{N},\frac{\pi}{N},\ldots
,\frac{(N-2)\pi}{N},\pi & \text{for even channel}%
\end{array}
\right. \label{alphak}%
\end{equation}
We can evaluate $\Gamma(r,N,\alpha_{k},\operatorname{e}^{\operatorname*{i}k})$
by applying the \emph{Proposition} directly to get%
\begin{equation}
\Gamma(r,N,\alpha_{k},\operatorname{e}^{\operatorname*{i}k})=\left(
-1\right)  ^{r}\left(  1-\frac{h^{2}}{J^{2}}\right)  ^{1/4}\left(  1-\frac
{2r}{N}\right)  =\Gamma(r,N,\alpha_{-k},\operatorname{e}^{-\operatorname*{i}%
k}),
\end{equation}
for large $r$ and $N$. So the asymptotic behavior of the correlation functions
is given by%
\begin{equation}
C_{r,N}^{x,x}\left(  |E_{q}\rangle\right)  =\left(  -1\right)  ^{r}\left(
1-\frac{h^{2}}{J^{2}}\right)  ^{1/4}\left(  1-\frac{2r}{N}\right)  ,
\end{equation}
which is independent of $q$. Thus, the correlation function is not sensitive
to the temperature if the temperature is low enough, i.e.%
\begin{equation}
C_{r,N}^{xx}(0\leq T\ll4h/k_{B})=\left(  -1\right)  ^{r}\left(  1-\frac{h^{2}%
}{J^{2}}\right)  ^{1/4}\left(  1-\frac{2r}{N}\right)  .\label{corall}%
\end{equation}

\section{II. \quad Experimental proposal}

\label{sec:exp} First, we propose how to construct a ring-geometric optical
lattice potential containing $N$ trapping wells. We only focus on the cases
where $N$ is an odd number, i.e. $N=2L+1(L=1,2,3,...)$. In $x$-$y$ plane we
can arrange $N$ beams of independent standing wave lasers which can be
obtained by frequency selection. Each standing wave will contributes an
optical potential along $\overrightarrow{k_{i}}$ direction which can be
expressed as $V_{x\text{-}y}\cos^{2}\left(  \overrightarrow{k_{i}}%
\cdot\overrightarrow{r_{i}}-\phi\right)  $ for i-th laser beam. $V_{x\text{-}%
y}$ is the strength of laser beams and $\phi$ is the phase shift. The angle
between two neighbor lasers is $2\pi/N$ [Fig.~5(a) in the main text]. Then, by
adopting $V_{x\text{-}y}$ and $\phi$, we find that we can always obtain a
circular lattice potential with $N$ traps in $x$-$y$ plane [Fig.~5(b) of the
main text]. As shown in Fig.~5(c) in the main text, in $z$ direction we apply
two independent standing wave lasers $V_{z1}\cos^{2}\left(  k_{z}z\right)  $
and $V_{z2}\cos^{2}\left(  2k_{z}z\right)  $ where one laser has twice wave
length of the second laser. The potential along $z$ direction has a double
well shape. By adopting the relative strengths $V_{z1}$\ and $V_{z2}$ of the
two lasers, we can neglect the hopping and interaction of cold atoms between
inter-double-well. Eventually, we obtain a periodical two-leg ladder potential
shown in Fig.~5(d) of the main text. In real experiment, there is additional
harmonic trapping potential $V_{trap}\left(  x^{2}+y^{2}\right)  $. The total
potential can be written as%
\begin{equation}
V_{potential}\left(  x,y,z\right)  =V_{trap}\left(  x^{2}+y^{2}\right)
+V_{z1}\cos^{2}\left(  k_{z}z\right)  +V_{z2}\cos^{2}\left(  2k_{z}z\right)
+V_{x\text{-}y}\sum_{i=1}^{N}\cos^{2}\left(  \overrightarrow{k_{i}}%
\cdot\overrightarrow{r_{i}}-\phi\right)  .\label{expvpotential}%
\end{equation}
\newline Second, we discuss the scheme to realize the transverse Ising model.
Let us consider loading into the ladders with cold atoms having two relevant
internal states that are denoted as pseudo-spin states $\lambda=\uparrow
,\downarrow$. The lattice potential experienced by cold atoms depends on which
of those two internal states are located. For sufficiently deep potential and
low temperatures, the system will be described by the following bosonic or
fermionic Hubbard model \cite{Duan},%
\begin{equation}
H_{Hub}=\sum_{j,\lambda,s}(-t_{\lambda})(a_{j\lambda,s}^{\dag}a_{(j+1)\lambda
,s}+\text{h.c.})+\sum_{j,\lambda}(-t_{\lambda})(a_{j\lambda,1}^{\dag
}a_{j\lambda,2}+\text{h.c.})+\frac{1}{2}\sum_{j,\lambda,s}U_{\lambda
}n_{j\lambda,s}(n_{j\lambda,s}-1)+\sum_{j,s}U_{\uparrow\downarrow}%
n_{j\uparrow,s}n_{j\downarrow,s},\label{exphhub}%
\end{equation}
where $s=1,2$ is the leg index. With the conditions of Mott insulator limit
$t_{\lambda}\ll U_{\lambda}$, $U_{\uparrow\downarrow}$ and half filling
$\langle n_{j\uparrow,s}\rangle+\langle n_{j\downarrow,s}\rangle\approx1$, the
low-energy Hamiltonian of Eq. (\ref{exphhub}) is mapped to the spin $XXZ$
model by second-order perturbation,%
\begin{equation}
H_{spin}=\sum_{j,s}\pm J_{\perp}(S_{j,s}^{x}S_{j+1,s}^{x}+S_{j,s}^{y}%
S_{j+1,s}^{y})+J_{z}S_{j,s}^{z}S_{j+1,s}^{z}+\sum_{j}\pm K_{\perp}(S_{j,1}%
^{x}S_{j,2}^{x}+S_{j,1}^{y}S_{j,2}^{y})+K_{z}S_{j,1}^{z}S_{j,2}^{z}%
,\label{exphspin}%
\end{equation}
where the pseudo-spin operator $\mathbf{S}=a^{\dag}\overrightarrow{\lambda
}a/2$, $\overrightarrow{\lambda}=(\lambda_{x},\lambda_{y},\lambda_{z})$ are
the Pauli matrices and $a^{\dag}=\left(  a_{\uparrow}^{\dag},a_{\downarrow
}^{\dag}\right)  $. The positive signs before $J_{\perp},K_{\perp}$ are for
fermionic atoms and negative signs for bosonic one. The interaction
coefficients for bosons are given by,
\begin{equation}
J_{\perp}=\frac{4t_{\uparrow}t_{\downarrow}}{U_{\uparrow\downarrow}}%
,J_{z}=\frac{2\left(  t_{\uparrow}^{2}+t_{\downarrow}^{2}\right)
}{U_{\uparrow\downarrow}}-\frac{t_{\uparrow}^{2}}{U_{\uparrow}}-\frac
{t_{\downarrow}^{2}}{U_{\downarrow}}\;,K_{\perp}=\frac{4t_{\uparrow}^{\prime
}t_{\downarrow}^{\prime}}{U_{\uparrow\downarrow}},K_{z}=\frac{2t_{\uparrow
}^{\prime2}+t_{\downarrow}^{\prime2}}{U_{\uparrow\downarrow}}-\frac
{t_{\uparrow}^{\prime2}}{U_{\uparrow}}-\frac{t_{\downarrow}^{\prime2}%
}{U_{\downarrow}}.\label{expjz}%
\end{equation}
For fermions, we only need to omit the last two terms in $J_{z}$ and $K_{z}$.
By modulating the intensity and the phase shift of the trapping laser beams
and by adjusting scattering length through Feshbach resonance, we can obtain a
desired Hamiltonian from Eq. (\ref{exphspin}),%
\begin{equation}
H_{s}=\sum_{j,s}J_{z}S_{j,s}^{z}S_{j+1,s}^{z}+\sum_{j}K_{{}}\vec{S}_{j,1}%
\cdot\vec{S}_{j,2}.
\end{equation}
The properties of this system are dominated by the pseudo-spin singlet
$\left\vert s\right\rangle _{j}=\left(  \left\vert \uparrow\downarrow
\right\rangle _{j}-\left\vert \downarrow\uparrow\right\rangle _{j}\right)
/\sqrt{2}$ and triplet $\left\vert t_{0}\right\rangle _{j}=\left(  \left\vert
\uparrow\downarrow\right\rangle _{j}+\left\vert \downarrow\uparrow
\right\rangle _{j}\right)  /\sqrt{2}$\ on the rung of the ladders in low
energy. At this time, the system can be mapped exactly to an effective FTIR
\cite{ChenQH},%
\begin{equation}
H_{T}=\sum_{j}2J_{z}\tilde{S}_{j}^{x}\tilde{S}_{j+1}^{x}-K_{{}}\tilde{S}%
_{j}^{z}.\label{expht}%
\end{equation}

\end{appendix}

\end{document}